# МНОГОЛУЧЕВОСТЬ В ЗЕРКАЛЬНЫХ РАДИОТЕЛЕСКОПАХ: ЧУВСТВИТЕЛЬНОСТЬ В ШИРОКОМ ПОЛЕ ОБЗОРА

Юпиков О.А.
Севастопольский национальный технический университет
99053, Севастополь, ул. Университетская 33, каф. радиотехники и телекоммуникаций
E-mail: lichne@gmail.com; тел. (050) 591-60-15

The given work is devoted to the modern developments in the field of radio astronomy instrumentation. In particular, the sensitivity of the multi-beam reflector radio telescope which is fed by phased array (PAF) is considered. Using PAF as reflector feed allows obtaining wide and continuous field of view (FOV) of the telescope. This has several advantages with compare to horn-cluster feeds which are described in this work. The sensitivity inside whole FOV was computed using three different beamforming schemes.

**Введение**

Традиционно в качестве облучателей зеркал рефлекторных антенн используются рупора, как правило, гофрированные. Однако, основная проблема современных радиотелескопов-интерферометров с рупорами — низкая скорость обзора неба. Чтобы получить максимальную разрешающую способность интерферометра, один цикл измерений длится 12 часов (пока Земля совершит пол оборота вокруг своей оси, т.о. получив самую длинную базу). За этот цикл просматривается только участок неба площадью равной площади поля обзора антенны. Так, например, у Вестерборгского радиотелескопа (*WSRT*, Нидерланды) ширина диаграммы направленности (ДН) на частоте 1,42 ГГц равна примерно 0,5 градуса. Несложно подсчитать, что при такой ширине ДН построение карты полной небесной сферы (площадью 4π кв. радиан) займет около 140 лет. На более высоких частотах это время будет еще больше.

Одним из способов увеличения скорости обзора неба является применение в качестве облучателей групп, или кластеров, рупоров в конфигурации «один рупор – один луч». Это позволяет одновременно проводить наблюдения нескольких участков неба. Были предложены различные варианты. Некоторые из них показаны на рис. 1 [1].

Однако, хотя использование кластеров из рупорных облучателей в конфигурации «один рупор – один луч» дает выигрыш в скорости обзора неба, оно имеет существенные недостатки: узкая рабочая полоса частот (для рупоров в составе кластеров она не превышает 30%) и низкая чувствительность в точках пересечения соседних лучей [1], из-за чего требуется несколько позиционирований зеркала, чтобы покрыть поле обзора.

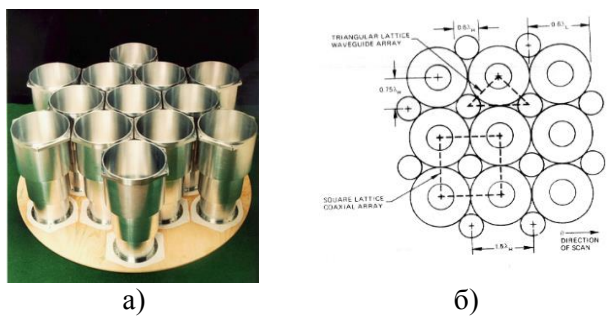

Рис. 1. Рупорные кластеры:
а) облучатель для *Parkes Radio Telescope*, (Δ*f*/*f* = 0,16); б) концепция двухполосного кластерного облучателя

Покрыть все поле обзора за одно позиционирование зеркала можно с использованием плотных антенных фокальных решеток (АФР), так как она позволяет формировать хорошо перекрываемые лучи (уровень в точках пересечения соседних лучей уменьшается не более чем на 3 дБ). Под понятием «плотная» антенная решетка понимается решетка с расстоянием между ее элементами на верхней частоте рабочего диапазона менее λ для рефлекторов с *F*/*D*>1 и менее 0,6λ для рефлекторов с *F/D*<0,5 [2]. Это необходимо, чтобы полностью семплировать поле в фокальной области.

АФР исследуются уже давно, начиная с 70х годов. Теоретически рассматривались такие вопросы, как оптимальное расстояние между элементами решетки и оптимальное отношение F/D зеркала, КИП, составляющие шумовой температуры системы, общие выражения для максимизации чувствительности системы с АФР, взаимное влияние элементов решетки друг на друга и вклад в шумовую температуру системы за счет этого, влияние искажения поверхности зеркала на параметры системы и другие вопросы. Однако анализа всей *многолучевой* системы в комплексе еще не было выполнено.

В настоящей работе приведенные выше результаты теоретических исследований объединяются в единую модель рефлекторной антенной системы с фокальной решеткой, и с помощью этой модели проводится анализ прототипа многолучевой системы *APERTIF* [3-4]. В докладе приводятся некоторые полученные результаты.

**Основная часть**

На рис. 2 показана функциональная схема анализируемой антенной системы. Она состоит из: 1) рефлектора, 2) плотной многоэлементной антенной решетки с устройством питания ее элементов, 3) малошумящих усилителей (МШУ), 4) цифрового формирователя ДН. С выхода каждого элемента решетки выходят как сигнал $e_m$, так и шум $c_m$.

В рассмотренной антенной системе формирователем одновременно формируются 37 лучей гексагональном в поле обзора площадью 8 кв. градусов (рис. 3). Окружностями показаны сечения лучей на уровне –3 дБ. Уровень –3 дБ задан в первом приближении, и может меняться в процессе оптимизации непрерывности поля обзора

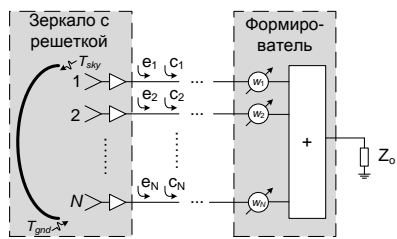 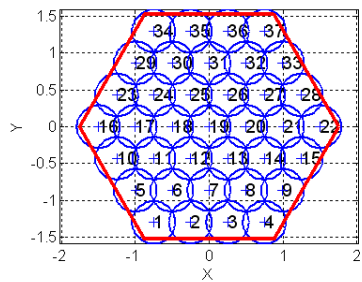

Рис. 2. Функциональная схема антенной системы  Рис. 3. Конфигурация лучей в поле обзора

Были рассмотрены три схемы формирования лучей: 1) метод согласования по полю [5]; 2) метод максимизации чувствительности [6]; 3) метод максимизации чувствительности с ограничениями по направлениям, известный также как *Linear Constrained Minimum Variance* (*LCMV*) [7].

**Метод согласования по полю (*CFM – conjugate field matching*)** — самый простой способ вычисления весовых коэффициентов для формирователя, при котором антенная система принимает максимум падающей на нее энергии, то есть ведется согласованный прием. Весовые коэффициенты **w**<sub>opt</sub> в этом методе равны комплексно-сопряженной величине сигнала на выходе соответствующих элементов решетки при приеме сигнала от источника в заданном направлении, то есть

$$\mathbf{w_{opt}} = \mathbf{e}^*\left(\theta_0, \varphi_0\right), \tag{1}$$

где $\mathbf{w_{opt}} = [w_1, w_2, \ldots, w_M]^T$ — весовые коэффициенты для приема падающей волны с заданной поляризацией; $(\theta_0, \varphi_0)$ — направление требуемого максимума ДН системы; $\mathbf{e}(\theta_0, \varphi_0)$ — сигнальный вектор при приеме падающей волны с направления $(\theta_0, \varphi_0)$.

Здесь и далее под сигнальным вектором **e** понимается набор значений выходного сигнала с каждого элемента решетки (канала приема), то есть $\mathbf{e} = [e_1, e_2, \ldots, e_M]^T$, где $M$ — количество антенных элементов в решетке, верхний индекс T означает транспонирование.

**Метод максимизации чувствительности (*MaxSNR*).** Оптимальные по критерию максимальной чувствительности весовые коэффициенты рассчитываются в соответствии со следующим выражением:

$$\mathbf{w_{opt}} = \mathbf{C}^{-1}\mathbf{e}(\theta_0, \varphi_0), \qquad (2)$$

где **C** — шумовая корреляционная матрица.

Элементы матрицы **C** являются коэффициентами корреляции между напряжениями на портах соответствующей пары элементов решетки как при приеме шумов окружения, так и учитывая внутренние шумы антенной системы и взаимную связь между элементами решетки. Более подробно о расчете этой матрицы см., например, [8].

**Метод максимизации чувствительности с ограничениями по направлениям (*LCMV*).** По данному методу формирования лучей весовые коэффициенты для формирователя рассчитываются следующим образом:

$$(\mathbf{w_{opt}})^H = \mathbf{g}^H \left[ \mathbf{G}^H \mathbf{C}^{-1} \mathbf{G} \right]^{-1} \mathbf{G}^H \mathbf{C}^{-1}, \qquad (3)$$

где **G** — матрица размером $M \times N_{dir}$, содержащая сигнальные вектора с $N_{dir}$ направлений ограничений ($M$ — количество элементов в решетке); **g** — вектор размерностью $N_{dir} \times 1$, элементы которого устанавливают уровень ДН луча в каждом из $N_{dir}$ направлений. Направления ограничений были выбраны в точках пересечения лучей.

**Моделирование антенной системы.**

На рис. 4 и 5 показаны фото и электродинамическая модель АФР.

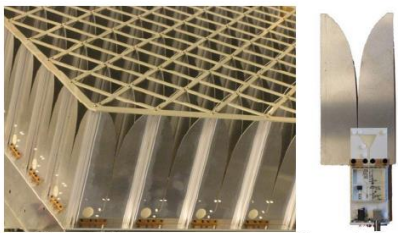 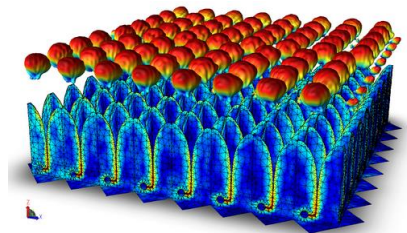

Рис. 4. Фото антенной решетки, установленной в фокусе *WSRT*, и один ее элемент

Рис. 5. Распределение токов в антенной решетке и ДН ее отдельных элементов

Используя созданную модель, были рассчитаны такие параметры радиотелескопа, как: а) весовые коэффициенты для элементов решетки; б) первичные и вторичные ДН для некоторых лучей; в) коэффициенты эффективности системы (КИП, коэффициент перехвата, коэффициенты амплитудного и фазового распределений в апертуре зеркала); г) шумовая температура и ее составляющие; д) КНД системы; е) чувствительность системы для нескольких лучей в поле обзора. Чувствительность в многолучевом поле обзора для трех рассмотренных схем формирования лучей, а также зависимость ее неравномерности от частоты показаны на рис. 6 и 7 соответственно.

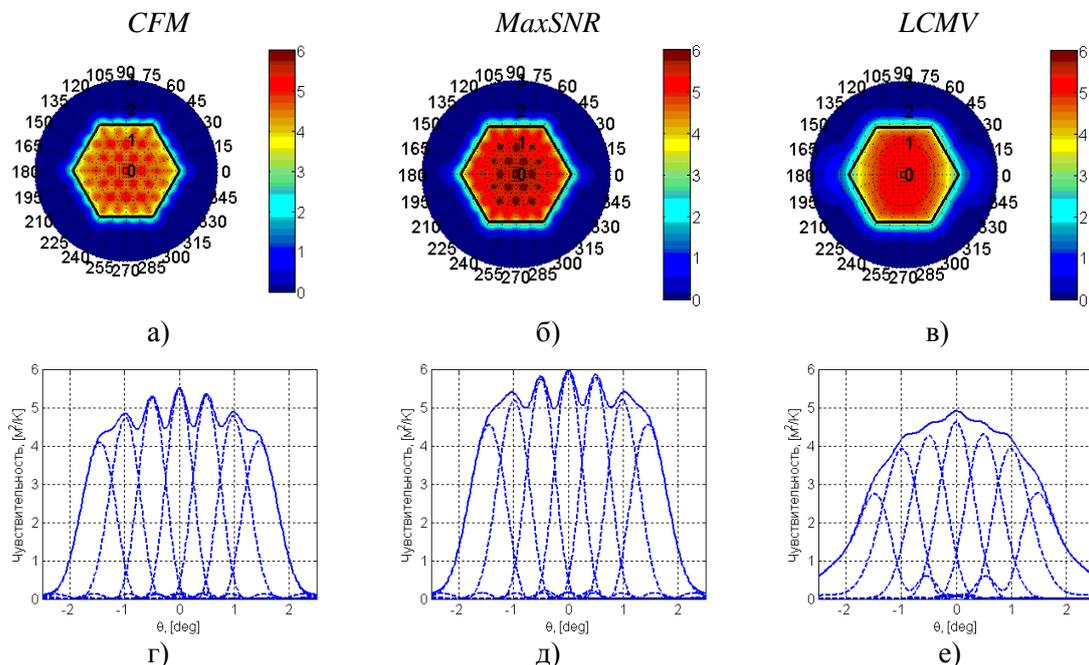

Рис. 6. Чувствительность в поле обзора антенной системы для трех схем формирования лучей

**Выводы**

Анализируя результаты, показанные на рис. 6 и 7, можно сделать следующие выводы.

1) схема *CFM* не дает максимальной чувствительности системы, хотя КНД при ней максимален, и обладает существенной неравномерностью в поле обзора, поэтому целесообразно применять другие схемы формирования лучей;

2) схема *LCMV* позволяет достичь компромисса между чувствительностью системы и ее неравномерностью внутри поля обзора;

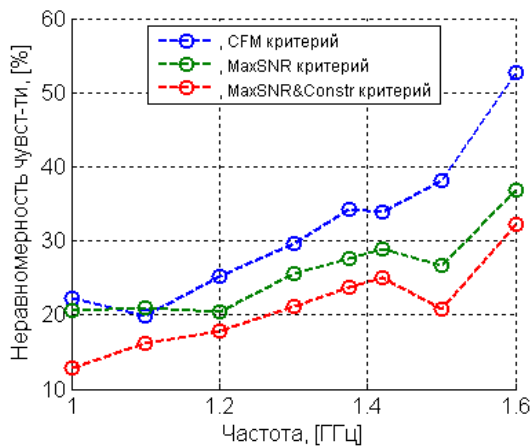

Рис. 7. Зависимость неравномерности чувствительности от частоты для трех схем

3) при использовании схемы *LCMV* чувствительность быстро падает при увеличении угла сканирования из-за искажения формы луча.

**Перечень ссылок.**

## Acknowledgement

This work has been supported by the Netherlands Institute for Radio Astronomy ASTRON, and conducted during Iupikov's visit to ASTRON during 2009 under the supervision of Drs. Ivashina, Maaskant and Cappellen.